\documentclass[aps,pra,twocolumn,showpacs,groupedaddress,amsmath]{revtex4-1}

\usepackage{bm}
\usepackage[english]{babel}
\usepackage{graphicx}
\usepackage{units}
\usepackage{hyperref}
\usepackage{ulem}
\usepackage{color}

\begin{document}

\title{Observation of Bose-enhanced photoassociation products}

\author{Alessio Ciamei}
\email[]{Sr2molecules@strontiumBEC.com}
\author{Alex Bayerle}
\author{Benjamin Pasquiou}
\author{Florian Schreck}
\affiliation{Van  der  Waals-Zeeman  Institute,  Institute  of  Physics,  University  of Amsterdam, Science  Park  904,  1098  XH  Amsterdam,  The  Netherlands}

\date{\today}


\begin{abstract}
We produce ${^{84}\mathrm{Sr}_2}$ molecules using Bose-enhanced Raman photoassociation. We apply the stimulated Raman adiabatic passage (STIRAP) technique on a Bose-Einstein condensate (BEC) to produce more than $8 \times 10^3$ ultracold molecules. This chemical reaction is only made possible because of the Bose enhancement of the optical transition dipole moment between the initial atomic state and an intermediate molecular state. We study the effect of Bose enhancement by measuring the transition Rabi frequency in a BEC and by comparing it with measurements for two atoms in sites of a Mott insulator. By breaking the dimers' bond and directly observing the separated atoms, we measure the molecular inelastic collision rate parameters. We discuss the possibility of applying Bose-enhanced STIRAP to convert a BEC of atoms into a BEC of molecules, and argue that the required efficiency for STIRAP is within experimental reach.
\end{abstract}

\maketitle

\section{Introduction}
\label{sec:Introduction}

The last decade has witnessed considerable progress in the study of chemical reactions at ultracold temperatures \cite{Jones2006RevPA, Chin2010RevFeshbach, Krems2008RevColdChemistry, Ospelkaus2010ChemReactKRb, Knoop2010ChemAtomDimer, Rui2016ChemReactAtomMol}. While direct cooling and trapping of molecules is being developped \cite{Barry2014MoleculeMOT, Kozyryev2017SysiphusCoolingMolecule}, so far ultracold chemistry relies on the formation of molecules from ultracold atoms. This process requires the existence of a coupling mechanism between a free-atom state and a bound molecular state. Feshbach resonances provide such coupling, and they have been exploited to produce molecules starting from a variety of ultracold bosonic and fermionic atom gases \cite{Xu2003FeshbachMolFromBEC, Regal2003FeshbachMolFromFermi}. Moreover, Feshbach resonances have been used to convert a Bose-Einstein condensate (BEC) of atoms into a Bose-condensed gas of Feshbach molecules, a molecular BEC (mBEC) \cite{Jochim2003molBEC, Greiner2003MolBECFromFermi}. However, suitable Feshbach resonances are only available in a limited number of systems, and they only allow the creation of molecules in high lying vibrational states, thus limiting experimental studies to a small class of chemical reactions. Another approach to the creation of a mBEC is Bose-enhanced stimulated Raman photoassociation (PA). This approach, which has been proposed in a number of theoretical works \cite{Javanainen1999PAonBEC, Drummond1998PASolitoninBEC, Julienne1998RamanPAmolInBEC}, involves the coupling by optical fields to an intermediate optically-excited molecular state. This technique requires the many-body Bose enhancement of the intrinsically weak coupling between the atomic state and the intermediate molecular state, and as such represents an example of \textit{superchemistry}, i.e. ``the coherent stimulation of chemical reactions via macroscopic occupation of a quantum state by a bosonic chemical species"\cite{Heinzen2000Superchemistry}. The main limitation in this optical scheme are the losses caused by spontaneous emission from the intermediate state. Stimulated Raman adiabatic passage (STIRAP) has been proposed as a method to minimize these losses \cite{Heinzen2000Superchemistry, Mackie2000BoseSTIRAPinPA, Drummond2002STIRAPmolBEC, Mackie2005CommentOnSTIRAPmolBEC, Drummond2005ReplyToCommentOnSTIRAPmolBEC, Bergmann2015RevSTIRAP}.

Until now, no direct observation of molecules produced by Bose-enhanced stimulated Raman photoassociation has been reported. Molecule creation through stimulated Raman PA has been demonstrated via atom loss spectroscopy of a BEC, where the stimulated transition rate was much smaller than the molecular lifetime \cite{Wynar2000molInBEC}. One-color PA of a BEC showing a non-classical association rate was reported in \cite{McKenzie2002PAsodiumInBEC}. Notably, the work of \cite{Yan2013RabiFlopBECMolviaPA} demonstrated reversible association by Rabi oscillations using a narrow one-color transition between atoms and molecules in a vibrational state born by an excited electronic potential. Moreover, the initial sample of \cite{Yan2013RabiFlopBECMolviaPA} is a BEC, which is coupled to a mBEC. Due to the excited state of the molecule, the yet undetermined lifetime of such a mBEC is expected to be short, and likely shorter than typical ground-state molecule lifetimes. A STIRAP exploiting the Bose enhancement of the dipole moment of the free-bound transition between the atomic state and the intermediate molecular state, would be a two-photon process capable of producing a mBEC of ground-state molecules. As an important step, two-color PA using a BEC was reported in \cite{Winkler2005AtomMolDarkStateInBEC}, showing the existence of atom-molecule dark states in a BEC, a necessary requirement for STIRAP.

In this paper, we report the optical production of ultracold ground-state ${^{84}\mathrm{Sr}_2}$ molecules starting from a BEC of Sr atoms, via a STIRAP pulse sequence using optical transitions in the vicinity of a narrow intercombination line. This result requires the Bose enhancement of the free-bound transition Rabi frequency, which we investigate by comparing the free-bound Rabi frequency measured in this paper to that measured for two isolated atoms in a Mott insulator sample \cite{Ciamei2017EffProductionSr2Mol}. By disassociating the dimers back into atoms, we directly observe that more than $8.1(0.7) \times 10^3$ ${\mathrm{Sr}_2}$ molecules are produced. As a first study of  the products of this chemical reaction, we measure the molecules inelastic collision rate parameters both with ${\mathrm{Sr}_2}$ molecules and Sr atoms. Our experimental demonstration indicates that STIRAP from an atomic BEC into a BEC of ground-state molecules could be observed, provided the STIRAP has a higher transfer efficiency, for which we suggest several methods.

\section{Experimental strategy}
\label{sec:ExperimentalStrategy}

\begin{figure}[tb]
\includegraphics[width=\columnwidth]{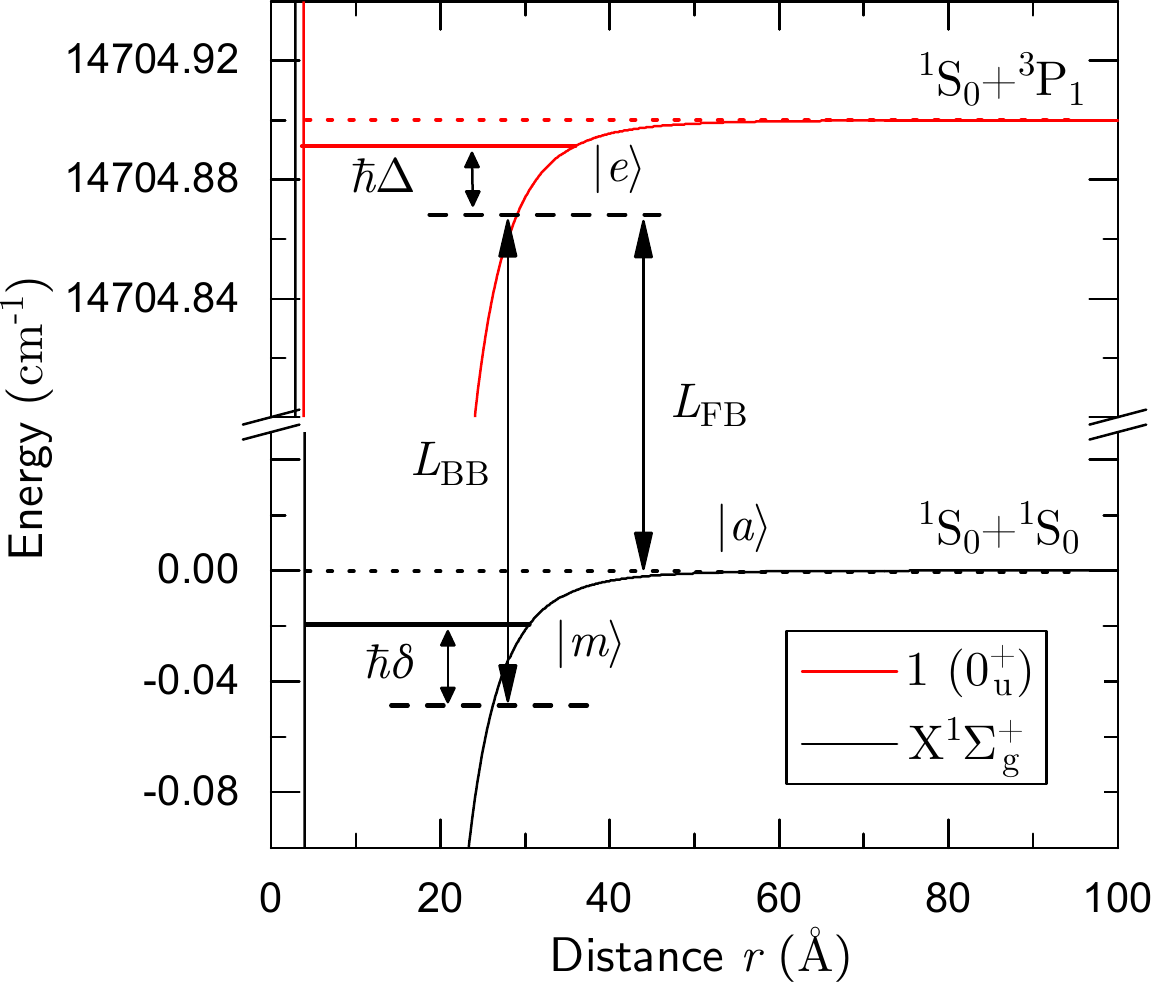}
\caption{\label{fig IntroPicture} (color online) ${^{84}\mathrm{Sr}_{2}}$ molecular potential for the electronic ground state ${X^{1}\Sigma^{+}_{g}}$ and the optically excited state ${1(0^{+}_{u})}$. The energy is referenced to the ground-state asymptote. The laser fields $L_{\mathrm{FB}}$ and $L_{\mathrm{BB}}$ are given, along with the one- and two-photon detunings $\Delta$ and $\delta$ for the $\Lambda$ scheme  $\{ \vert a \rangle , \vert e \rangle, \vert m \rangle \}$ used for STIRAP.}
\end{figure}

We apply a STIRAP sequence to convert Sr atoms in a BEC into molecules, using the $\Lambda$ scheme that was adopted in our previous work \cite{Stellmer2012Sr2Mol}. Figure~\ref{fig IntroPicture} shows the relevant potential energy curves of ${^{84}\mathrm{Sr}_{2}}$ in the Born-Oppenheimer approximation. The initial sample is a BEC composed of atoms in the electronic ground-state ${^1\mathrm{S}_0}$, and we label $\vert a \rangle$ the state associated with a pair of such atoms. The produced molecules populate the bound state $\vert m \rangle$, with binding energy $\Delta E_{m} = h \times 644.7372(2) \unit{MHz}$ (with $h$ being the Planck constant), corresponding to the second to last vibrational state $\nu=-2$ of the potential ${X^{1}\Sigma^{+}_{g}}$, which asymptotically correlates to two Sr atoms in ${^1\mathrm{S}_0}$. We couple these two states via a third state $\vert e \rangle$, a molecular bound state with binding energy $\Delta E_{e}= h \times 228.38(1)\unit{MHz}$, corresponding to the vibrational state $\nu=-3$ of the potential ${1(0^{+}_{u})}$, which correlates to one atom in ${^1\mathrm{S}_0}$ and one in the optically excited electronic state ${^3\mathrm{P}_1}$. The states $\vert a \rangle$ and $\vert e \rangle$ are coupled with Rabi frequency $\Omega_{\mathrm{FB}}$ by the free-bound laser $L_{\mathrm{FB}}$ and $\vert e \rangle$ is coupled to $\vert m \rangle$ with Rabi frequency $\Omega_{\mathrm{BB}}$ by the bound-bound laser $L_{\mathrm{BB}}$.

The STIRAP $\Lambda$ scheme is coupled to the environment mainly through spontaneous emission from $\vert e \rangle$. The STIRAP sequence exploits the presence of a dark state, i.e. an eigenstate of the system orthogonal to $\vert e \rangle$, which can be adiabatically moved from $\vert a \rangle$ to $\vert m \rangle$ over the sequence time $T$, thus providing near-unit efficiency of atom-molecule conversion. The rotation of the dark state is controlled by the parameter $\Omega_{\mathrm{FB}}/\Omega_{\mathrm{BB}}$, which should be $\Omega_{\mathrm{FB}}/\Omega_{\mathrm{BB}}\ll 1$ at the initial time and $\Omega_{\mathrm{FB}}/\Omega_{\mathrm{BB}}\gg 1$ at time $T$. In a STIRAP sequence, the parameter $\Omega_{\mathrm{FB}}/\Omega_{\mathrm{BB}}$ is tuned by the temporal intensity profile of $L_{\mathrm{FB}}$ and $L_{\mathrm{BB}}$.

\begin{figure}[tb]
\includegraphics[width=\columnwidth]{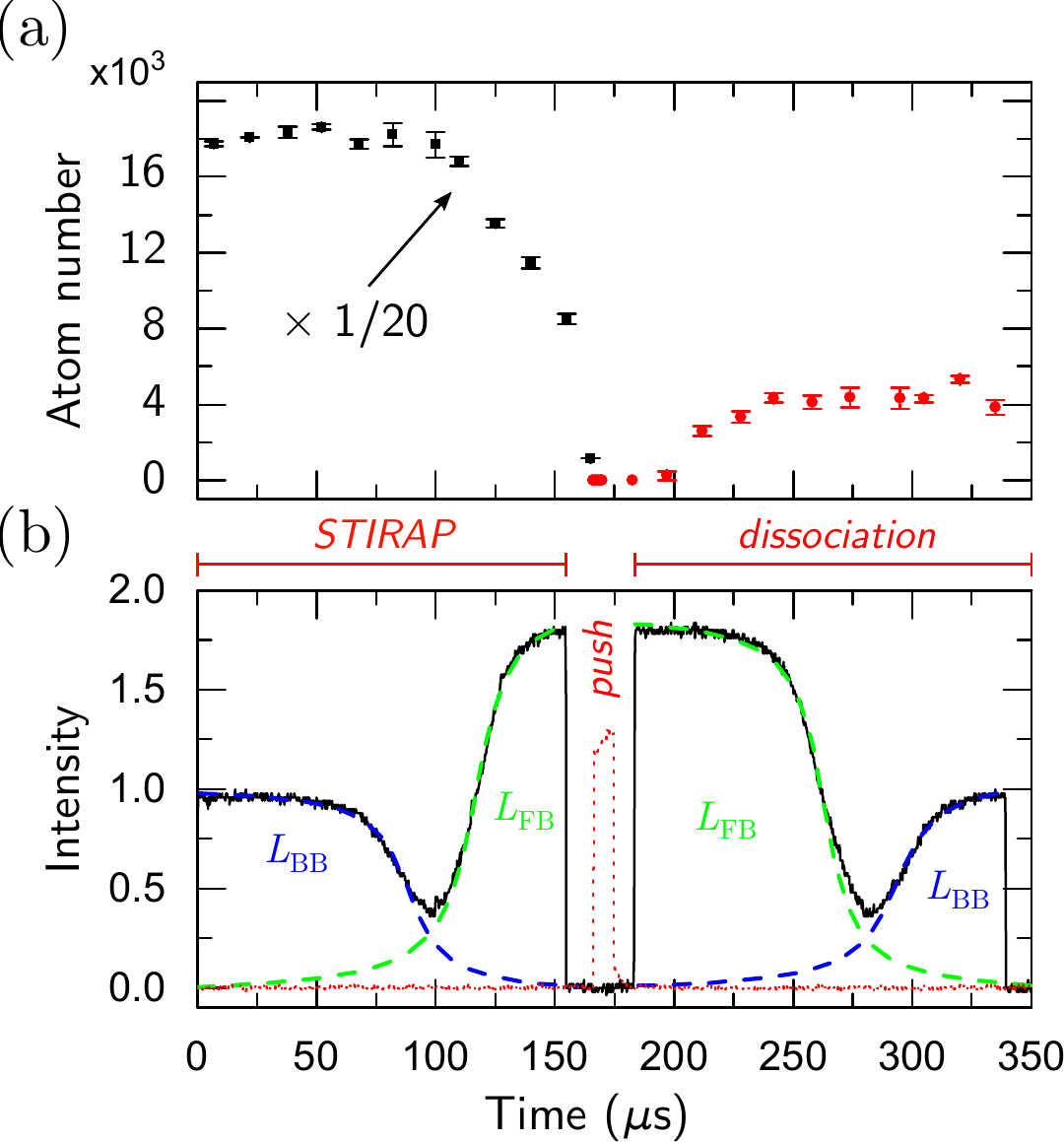}
\caption{\label{fig STIRAPBEC} (color online) Typical time behaviour of the number of atoms, during a sequence composed of STIRAP ($\vert a \rangle \rightarrow \vert m \rangle$) --- push pulse --- dissociation ($\vert m \rangle \rightarrow \vert a \rangle$). (a) Atom number during the STIRAP (black squares) and dissociation (red circles). The atom number during STIRAP has been scaled by a factor $1/20$ for clarity. The only atoms present after the dissociation transfer come from dissociated molecules. (b) Combined photoassociation laser intensities on the atomic sample (continuous black line). The dashed lines represent the contribution of $L_{\mathrm{FB}}$ and $L_{\mathrm{BB}}$. The dotted red line gives the intensity of the push pulse between STIRAP and dissociation.}
\end{figure}

In order to observe the molecules we produce, we disassociate them back into atoms. We first produce $\mathrm{Sr}_2$ by a STIRAP sequence that induces the transfer $\vert a \rangle \rightarrow \vert m \rangle$. At the end of the STIRAP, we push away all remaining atoms by a pulse of light resonant with the ${^1\mathrm{S}_0} - {^1\mathrm{P}_1}$ transition, therefore leaving only molecules in the trap. The molecules remain unaffected by this pulse, because their binding energy is much bigger than the transition linewidth $\Gamma_{^1\mathrm{P}_1} \simeq 2\pi \times 30 \unit{MHz}$. The reverse transfer $\vert m \rangle \rightarrow \vert a \rangle$ is obtained by the time-mirrored sequence of the laser beams intensity ramps used for STIRAP (see Fig.~\ref{fig STIRAPBEC}b). We image the atoms resulting from dissociated molecules by using absorption imaging on the ${^1\mathrm{S}_0} - {^1\mathrm{P}_1}$ transition. Despite the time-symmetry of the laser intensity ramps, the dissociation sequence is not related to a STIRAP process. Indeed, if population transfer is mostly due to Bose-enhanced two-photon processes, the dissociation efficiency can be significantly lower than the association efficiency, and the enhancement depends on the population of state $\vert a \rangle$ surviving the push beam. We recover atoms by the dissociation sequence owing to both two-photon stimulated emission and spontaneous emission from $\vert e \rangle$. Therefore, the stated numbers of produced molecules are a conservative lower bound, which assumes an unrealistic $100\unit{\%}$ dissociation efficiency.

We produce the initial BEC of ${^{84}\mathrm{Sr}}$ as in \cite{Stellmer2013QDegSr}. The BEC contains typically $N \approx 3.0\times 10^5$ atoms with a peak density of $n_{\mathrm{peak}}= 1.9(1) \times 10^{14} \unit{cm^{-3}}$. The trap frequencies are $\omega_{x}= 2\pi \times 34 \unit{Hz}$, $\omega_{y}= 2\pi \times 22\unit{Hz}$ and $\omega_{z}= 2\pi \times 430 \unit{Hz}$, where the $z$-axis is vertical. The Thomas-Fermi radii are $R_{x}= 20 \unit{\mu m}$, $R_{y}= 31 \unit{\mu m}$ and $R_{z}= 1.6 \unit{\mu m}$, and the chemical potential is $\mu = 90(3) \unit{nK}$.

The beam containing both laser fields, $L_{\mathrm{FB}}$ and $L_{\mathrm{BB}}$, used for PA is horizontal and has a waist of $113(2) \unit{\mu m}$. The polarization is linear and parallel to a vertically oriented guiding magnetic field of $5.30(5) \unit{G}$ and thus only $\pi$ transitions can be addressed. This field splits the Zeeman levels of state $\vert e \rangle$ by $2\pi \times 1.65(1) \unit{MHz}$. Both $L_{\mathrm{FB}}$ and $L_{\mathrm{BB}}$ are derived from two injection-locked slave lasers seeded by the same master oscillator with a linewidth of less than $2\pi \times 3 \unit{kHz}$. The frequencies of these laser fields are tuned by acousto-optical modulators and the beams are combined into a single-mode fiber with the same polarization, so that the main difference on the atomic cloud are their intensity and frequency. This setup ensures good coherence between the two fields, whose beat note must match the Raman condition, i.e. the frequency difference must be equal to the binding energy of state $\vert m \rangle$.

\section{Theory}
\label{sec:TheoreticalModel}

The evolution of the population during the photoassociation sequence starting from a BEC is modelled by the set of equations \cite{Winkler2005AtomMolDarkStateInBEC, Mackie2000BoseSTIRAPinPA}

\begin{equation}
\label{eq BEC Model}
\begin{cases}
& i\,\dot{a} = -i\frac{\gamma_{a}}{2}\,a - \Omega_{\mathrm{FB}}\,a^*\,e\, ,\\
& i\,\dot{e} = -\frac{1}{2}(\Omega_{\mathrm{FB}}\,a^2 + \Omega_{\mathrm{BB}}\,m )+ (\Delta -i\frac{\gamma_{e}}{2}) e\, ,\\
& i\,\dot{m} = -\frac{1}{2}\Omega_{\mathrm{BB}}\,e + (\delta -i\frac{\gamma_{m}}{2}) m\, ,\\
\end{cases}
\end{equation}

\noindent where $a=a(t)$, $e=e(t)$ and $m=m(t)$ are the amplitudes corresponding respectively to the atomic condensate field, the excited molecular field and the ground-state molecular field in the semi-classical approximation. Losses are described by the parameters $\gamma_{a}$, $\gamma_{e}$ and $\gamma_{m}$, which originate from the coupling of the three level system to the environment and which make the effective Hamiltonian non-Hermitian. In the absence of losses, the field amplitudes are normalized according to $\vert a \vert ^2 + 2 \vert e \vert^2+ 2 \vert m \vert^2 = 1$. The free-bound Rabi frequency $\Omega_{\mathrm{FB}}$ between the atomic field and the excited molecular field is given by $\Omega_{\mathrm{FB}}= d^{\mathrm{BEC}}_{\mathrm{FB}}\times E$, where $E$ is the electric field amplitude and $d^{\mathrm{BEC}}_{\mathrm{FB}}$ is the Bose-enhanced transition dipole moment between the two fields. This quantity can be written as $d^{\mathrm{BEC}}_{\mathrm{FB}}=d^{\mathrm{Bare}}_{\mathrm{FB}}\times \sqrt{N}$, where $d^{\mathrm{Bare}}_{\mathrm{FB}}$ is the bare transition dipole moment calculated for a single atom pair of the condensate, and $N$ is the number of atoms in the condensate \cite{Mackie2000BoseSTIRAPinPA}. Finally, the terms $\Delta$ and $\delta$ are respectively the one-photon and two-photon detunings shown in Fig.~\ref{fig IntroPicture}.

A criterion for a STIRAP  transfer with near unit efficiency is to maintain the adiabaticity of the evolution of the dark state throughout the sequence. As discussed in \cite{Vitanov1997StirapPopTransferTheory, Ciamei2017EffProductionSr2Mol}, this translates into the constraint $\alpha = \tilde{\gamma} T \gg 1$, with $ \tilde{\gamma} =\Omega_{\mathrm{FB,BB}}^2 / \gamma_{e}$. For realistic experimental conditions, the sequence time $T$ is limited by the lifetimes of states $\vert a \rangle$ and $\vert m \rangle$, and the finite lifetime of the dark state. This puts a lower bound on the two Rabi frequencies $\Omega_{\mathrm{FB,BB}}$. A good choice of the pair of states $\vert e \rangle$ and $\vert m \rangle$ can  ensure a suitably strong bound-bound Rabi frequency $\Omega_{\mathrm{BB}}$. On the contrary, the free-bound transition is intrinsically weak, due to the very small overlap between the wavefunction of a pair of atoms, whose characteristic length is determined by the trap, and the wavefunction of a single molecule, whose characteristic length is determined by the --- much smaller --- Condon point. Typically, for equal laser intensity, the free-bound Rabi frequency is several orders of magnitude smaller than the bound-bound Rabi frequency, which would make STIRAP impossible for realistic experimental condittions. Since a BEC can contain on the order of 10$^3$ - 10$^8$ atoms \cite{vanderStam2007BigNaBEC}, it has been predicted that the Bose-enhanced $\Omega_{\mathrm{FB}}$ can lead to a near-unit STIRAP efficiency \cite{Mackie2000BoseSTIRAPinPA}.

\section{Bose-enhanced Rabi frequency}
\label{sec:BoseEnhancedRabiFrequency}

We experimentally determine the free-bound Rabi frequency $\Omega_{\mathrm{FB}}$ for a BEC. This determination requires the knowledge of the natural linewidth of the excited state $\Gamma_e$, which we assume is  in our case the only contribution to the losses described by $\gamma_e$, i.e. $\gamma_e = \Gamma_e$. In order to measure $\Gamma_e$, we shine $L_{\mathrm{FB}}$ on the BEC and record both the decay of the atom number over time and the spectral width of the loss signal. We make sure to use low intensities of the PA light in order to make light shifts irrelevant. In order to extract $\Gamma_e$ from these data sets, we simplify our model by adiabatically eliminating the variable $e$ from eq.~(\ref{eq BEC Model}), imposing $\dot{e}=0$. The resulting equation for the atomic amplitude is $\dot{a}=-\frac{\Omega_{\mathrm{FB}}^2}{\Gamma_e}\,a^3$, which can be written in terms of the BEC atom number $\dot{N}=-K\, N^2$. We now first fit the data sets featuring the atom number decay with the function $N(t)=\frac{N_{0}}{1+K N_{0} t}$ and obtain $\frac{\Omega_{\mathrm{FB}}^2}{\Gamma_e}=\frac{K N_{0}}{2}$. Second, we fit the experimental spectral widths with the theoretical ones predicted by the model for a known value of ${\Omega_{\mathrm{FB}}^2}/{\Gamma_e}$, and thus obtain $\Gamma_e$. We derive the free-bound natural linewidth $\Gamma_e=2 \pi \times 19.2(2.4) \unit{kHz}$, which is consistent with measurements using a Mott insulator sample \cite{Ciamei2017EffProductionSr2Mol} and with values measured for ${^{88}\mathrm{Sr}_{2}}$ molecules \cite{Reinaudi2012Sr2Mol}. The error in $\Gamma_e$ is dominated by the error in the measured width in one-color PA spectra. 

In order to determine $\Omega_{\mathrm{FB}}$, we next measure time-decay curves for a wide range of $L_{\mathrm{FB}}$ intensities, at the fixed density $n_{\mathrm{peak}}= 1.9 \times 10^{14} \unit{cm^{-3}}$ of the BEC. The experimental curve for $\Omega_{\mathrm{FB}}$ as function of $\sqrt{I_{\mathrm{FB}}}$ presents deviations from linearity due to light shifts in the high power regime, and to inelastic processes at low Rabi frequencies and long pulse times. We derive a lower limit for the free-bound Rabi frequency of $\Omega_{\mathrm{FB}} = 2 \pi \times 3.6(6) \unit{kHz / \sqrt{W/cm^{2}}}$, from the experimental data sets corresponding to high $L_{\mathrm{FB}}$ intensities, where the effect of inelastic processes vanishes but the light shifts remain negligible.

\begin{figure}[tb]
\includegraphics[width=\columnwidth]{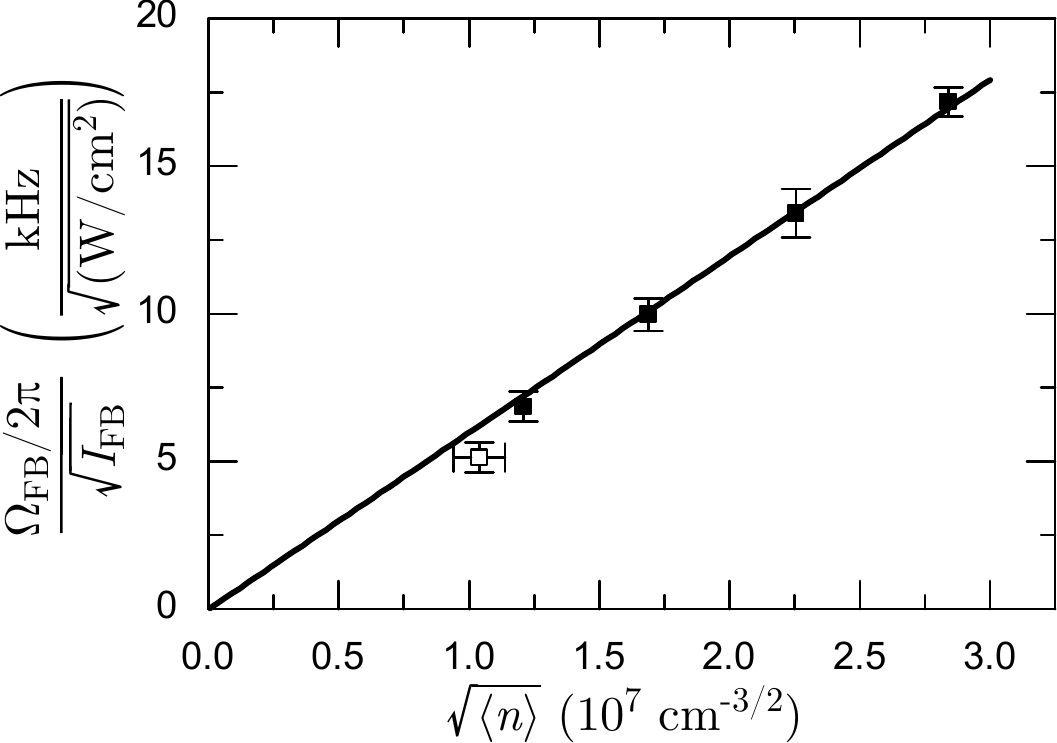}
\caption{\label{fig FBOmegaVSExtConf} Free-bound Rabi frequency as a function of $ \sqrt{\langle n \rangle}$, where $\langle n \rangle$ is the spatially averaged atomic density, measured in a BEC (empty square) and in a MI (filled squares). The line is a fit of the form $\Omega_{\mathrm{FB}}= c_1 \sqrt{\langle n \rangle}$ to all data points.}
\end{figure}

In order to demonstrate the Bose enhancement of the dipole moment of the free-bound transition, we compare our measurement of $\Omega_{\mathrm{FB}}$ with the one performed in a Mott insulator (MI) sample of doubly-occupied sites \cite{Ciamei2017EffProductionSr2Mol}. In Fig.~\ref{fig FBOmegaVSExtConf} we show the free-bound Rabi frequency as a function of $\sqrt{\langle n \rangle}$, where $\langle \cdot \rangle$ represents the spatial average. For the measurement presented in this article, $n$ is the density of the BEC calculated under the Thomas-Fermi approximation. For the measurements from \cite{Ciamei2017EffProductionSr2Mol}, $n$ is the on-site density of a single atom. Moreover, the PA process is described by two different models in the case of a BEC (see eq.~(\ref{eq BEC Model})) and in the case of a MI (see eq. (1) in \cite{Ciamei2017EffProductionSr2Mol}), so we need to be careful when comparing the free-bound Rabi frequencies. For this plot, we define $\Omega_{\mathrm{FB}}$ by considering the atom decay rate $W_{a}$ evaluated experimentally at the beginning of the PA pulse. As shown previously, in the BEC case the solution to the model is $\dot{a}=-\frac{\Omega_{\mathrm{FB}}^2}{\Gamma_e}\,a^3$, and therefore $W_{a}=-2 {\Omega^{\mathrm{BEC}}_{\mathrm{FB}}}^2/\Gamma_e$. In the MI case, the solution is $\dot{a}=-\frac{\Omega_{\mathrm{FB}}^2}{2 \Gamma_e}\,a$, which gives the rate $W_{a}=- {\Omega^{\mathrm{MI}}_{\mathrm{FB}}}^2/\Gamma_e$. This difference is a consequence of the different normalization used in the two models. In Fig.~\ref{fig FBOmegaVSExtConf}, we plot $\Omega_{\mathrm{FB}} = \sqrt{2} \Omega^{\mathrm{BEC}}_{\mathrm{FB}}$ for the BEC case (empty square).

To explain why our data demonstrates Bose enhancement, we first consider the BEC case and the MI case as two separate problems. We call $n_\mathrm{BEC}$ the BEC density and we call $n_\mathrm{MI}$ the on-site density in the MI. As discussed in detail in \cite{Ciamei2017EffProductionSr2Mol}, the free-bound Rabi frequency for two atoms in a lattice site of the MI is $\Omega^{{\mathrm{MI}}}_{\mathrm{FB}}=c \sqrt{\langle n_\mathrm{MI} \rangle}$, where $c$ is a constant that depends on the chosen free-bound PA line but not on the atomic density. We assume that this relation can be applied to the case of two atoms in a BEC, provided that we use $n_\mathrm{MI}=n_\mathrm{BEC}/N$, where $N$ is the BEC atom number. The Rabi frequency for two atoms, i.e. the bare Rabi frequency, is then given by $\Omega^{{\mathrm{Bare}}}_{\mathrm{FB}}=c \sqrt{\langle n_\mathrm{BEC}/N \rangle}$, which, for our experimental conditions, is roughly three orders of magnitude smaller than $\Omega^{{\mathrm{MI}}}_{\mathrm{FB}}$. However in presence of Bose enhancement, which features pair-wise constructive interference of the transition dipole moments, the bare Rabi frequency is multiplied by $\sqrt{N}$, and thus $\Omega^{{\mathrm{BEC}}}_{\mathrm{FB}}= \sqrt{N} \, \Omega^{{\mathrm{Bare}}}_{\mathrm{FB}} = c \sqrt{\langle n_\mathrm{BEC} \rangle}$. As a consequence, we fit the BEC data together with the MI data, using a function with the form $\Omega_{\mathrm{FB}}= c_1 \sqrt{\langle n \rangle}$. The fitted function is shown in Fig.~\ref{fig FBOmegaVSExtConf}. The fit is successful, which demonstrates the Bose enhancement in the BEC case, and we derive $\Omega_{\mathrm{FB}}= 2 \pi \times 5.9(2) \, 10^{-7} \unit{kHz \, cm^{3/2} / \sqrt{W/cm^{2}}}$.

\section{STIRAP parameters}
\label{sec:STIRAPParameters}

Before producing molecules by STIRAP, we first measure the other parameters relevant to the STIRAP transfer efficiency, which are necessary inputs to solve the model of eq.~(\ref{eq BEC Model}) numerically. The results of our measurements together with the theoretical estimates of these parameters are compiled in Tab.~\ref{tab ParametersBEC}.

\begin{table}[tb] 
\caption{Parameters of the chosen $\Lambda$ scheme used for molecule creation from a BEC, measured experimentally and estimated theoretically. Symbols are defined in the text and illustrated in Fig.~\ref{fig IntroPicture}. The theoretical values of $\Omega_{\mathrm{FB}}$ and $\Omega_{\mathrm{BB}}$ are obtained by using mapped grid methods \cite{Willner2004MapGridMethods} and the WKB approximation with the potentials given in \cite{Stein2010HeatPipePotSr2, Borkowski2014MassScalingSr2}.} 
\centering      
\begin{tabular}{c c c c}                    
\\
Param. & Units & Experiment & Theory \\ [0.5ex]
\hline  \\                
$\Omega_{\mathrm{FB}}$ & $ \unit{\frac{kHz \, cm^{3/2}}{\sqrt{W/cm^{2}}}}$ & $2\pi \times 5.9(2) \, 10^{-7}$ &  $2\pi \times 3.5 \, 10^{-7}$ \\ 
$\Omega_{\mathrm{BB}}$ & $ \unit{\frac{kHz}{\sqrt{W/cm^{2}}}}$ & $2\pi \times 234(5)$ & $2\pi \times 660 $\\ 
$\Gamma_e$ & $\unit{kHz}$ & $2\pi \times 19.2(2.4)$ & $>2\pi \times 14.8$\\   
$\Delta_{\mathrm{FB}}$ & $ \unit{\frac{kHz}{W/cm^{2}}}$ & $+ 2\pi \times 21(1)$ & $+ 2\pi \times 20.0$\\ 
$\delta_{\mathrm{FB}}$ & $ \unit{\frac{kHz}{W/cm^{2}}}$ & $+ 2\pi \times 17.0(2)$ & $+ 2\pi \times 20.0$\\
$\Delta_{\mathrm{DT}}$ & $ \unit{\frac{Hz}{W/cm^{2}}}$ & $+ 2\pi \times 2.61(8)$ & -\\
$\tau_{\mathrm{Dark}}$ & $\unit{ms}$ & 2.6(3) & -\\ [1ex]
\hline     
\end{tabular} 
\label{tab ParametersBEC}  
\end{table}

The bound-bound Rabi frequency $\Omega_{\mathrm{BB}}$ can be directly measured through loss spectroscopy, by probing the Autler-Townes splitting induced by $\Omega_{\mathrm{BB}}$ with the free-bound laser $L_{\mathrm{FB}}$ \cite{Autler1955AutlerTownesSplitting}. We measure $\Omega_{\mathrm{BB}}= 2\pi \times 234(5) \unit{kHz /\sqrt{W/cm^{2}}}$, where the measurement error is the standard deviation resulting from the fit.

In our approach, both the one-photon detuning $\Delta$ and the two-photon detuning $\delta$ depend on light shifts induced on the states of our $\Lambda$ scheme. The free-bound laser $L_{\mathrm{FB}}$, whose intensity varies during STIRAP, induces a time- and space-dependent shift $\Delta_{\mathrm{FB}}$ on the free-bound transition. It also induces a time- and space-dependent shift $\delta_{\mathrm{FB}}$ on the binding energy of state $\vert m \rangle$. The light shifts from the weak $L_{\mathrm{BB}}$ laser are always negligible. We measure $\Delta_{\mathrm{FB}}$ and $\delta_{\mathrm{FB}}$ by performing one- and two-color spectroscopy for several intensities of $L_{\mathrm{FB}}$, and obtain $\Delta_{\mathrm{FB}}= 2\pi \times 21(1) \unit{kHz / \sqrt{W / cm^{2}}}$ and $\delta_{\mathrm{FB}}=2\pi \times 17.0(2) \unit{kHz / \sqrt{W / cm^{2}}}$. Both shifts are dominated by the shifts of state $\vert a \rangle$ induced by the off-resonant transition ${^1\mathrm{S}_0} - {^3\mathrm{P}_1} \, (m_J=0)$, as discussed in \cite{Ciamei2017EffProductionSr2Mol}. More precisely $\Delta_{\mathrm{FB}} \approx \delta_{\mathrm{FB}} \approx \hbar \Omega_{\mathrm{FF}}^{2} /2\Delta E_{e}$, where $\hbar$ is the reduced Planck constant and $\Omega_{\mathrm{FF}}$ is the Rabi frequency induced on the atomic transition ${^1\mathrm{S}_0} - {^3\mathrm{P}_1}$ by $L_{\mathrm{FB}}$.

The DT induces space-dependent but time-independent light shifts $\Delta_{\mathrm{DT}}$ and $\delta_{\mathrm{DT}}$, respectively on the free-bound transition and on the binding energy of state $\vert m \rangle$. We measure $\Delta_{\mathrm{DT}} = 2\pi \times 2.61(8) \unit{Hz / (W / cm^{2})}$, while $\delta_{\mathrm{DT}}$ is here negligible \cite{Ciamei2017EffProductionSr2Mol}. These time-independent shifts are compensated for by adapting the frequencies of the PA lasers. The remaining inhomogeneous energy spread originating from the finite size of the sample and the spatial profile of the trapping potential is too small to influence the STIRAP transfer efficiency and can be neglected. On the contrary, the time-dependent shifts $\Delta_{\mathrm{FB}}$ and $\delta_{\mathrm{FB}}$ are not compensated for at all times, and can limit the efficiency.     

The lifetime of the dark-state superposition $\tau_{\mathrm{Dark}}$ engineered during a STIRAP must be longer than the STIRAP sequence time in order to achieve a high transfer efficiency. We measure this lifetime by shining both $L_{\mathrm{FB}}$ and $L_{\mathrm{BB}}$ on the BEC, with $\Delta=\delta=0$, and with $\Omega_{\mathrm{FB}}= 2\pi \times 3.5 \unit{kHz}$ and $\Omega_{\mathrm{BB}}= 2\pi \times 10 \unit{kHz}$. These Rabi frequencies are chosen in order for the dark state to have a significant overlap with the initial atomic state. We observe a first fast exponential decay of the atom number and a second slower one, with $1/e$ time constants of $100(20) \unit{\mu s}$ and $2.6(3) \unit{ms}$ respectively. The former time constant corresponds to scattering on the free-bound line, while the latter corresponds to the lifetime $\tau_{\mathrm{Dark}}$. One possible origin for this finite lifetime is the spatially varying mean-field shift across the BEC. We estimate this shift to be $2 \pi \times 0.37(5) \unit{kHz}$, which gives a dephasing time of $2.7(4) \unit{ms}$. 

As shown in measurements presented in the following paragraphs, the molecule lifetime is long compared to the STIRAP pulse duration, so we can neglect the effect of $\gamma_{m}$. Finally, the atomic loss term $\gamma_{a}$ is dominated by off-resonant scattering on the atomic ${^1\mathrm{S}_0} - {^3\mathrm{P}_1}$ line and can also be neglected.

\section{Molecule production}
\label{sec:MoleculeProduction}

We now apply STIRAP to a BEC and study its effects. We optimize the $\vert a \rangle \rightarrow \vert m \rangle$ transfer efficiency by varying the parameters $\Omega_{\mathrm{FB}}$, $\Omega_{\mathrm{BB}}$, $T_{\mathrm{pulse}}$ and $n_{\mathrm{peak}}$ while aiming for the maximum atom number after a STIRAP sequence followed by the push pulse and the dissociation transfer $\vert m \rangle \rightarrow \vert a \rangle$, see Fig.~\ref{fig STIRAPBEC}. We obtain the best efficiency for $\Omega_{\mathrm{FB}} = 2\pi \times 6.0(4) \unit{kHz}$, $\Omega_{\mathrm{BB}} = 2\pi \times 610(10) \unit{kHz}$, $T_{\mathrm{pulse}} = 150 \unit{\mu s}$, and $n_{\mathrm{peak}} = 0.44(2) \times 10^{14} \unit{cm^{-3}}$. For this optimization, we also vary the frequencies of $L_{\mathrm{FB},\mathrm{BB}}$, but we keep them constant during the whole STIRAP+dissociation sequence. The maximum number of atoms having successfully undergone the $\vert a \rangle \rightarrow \vert m \rangle$ then $\vert m \rangle \rightarrow \vert a \rangle$ sequence is $16.2(1.4) \times 10^3$. Assuming a hypothetical $100 \unit{\%}$ efficiency for the $\vert m \rangle \rightarrow \vert a \rangle$ transfer, we derive a lower limit for the number of molecules produced of $N_{\mathrm{mol,MIN}} = 8.1(0.7) \times 10^3$. Under this assumption, the lower limit for the STIRAP efficiency is $\eta_{\mathrm{MIN}} = 5.1(6) \unit{\%}$, which is orders of magnitude higher than the minute $3.1(8)\times 10^{-7} \unit{\%}$ efficiency that is expected without Bose-enhancement of $\Omega_{\mathrm{FB}}$. For this reason, the non-zero number of atoms having successfully undergone the $\vert a \rangle \rightarrow \vert m \rangle$ then $\vert m \rangle \rightarrow \vert a \rangle$ sequence is proof of the Bose enhancement of the free-bound transition dipole moment. For the experimentally optimized parameters, our numerical model using eq.~(\ref{eq BEC Model}) predicts an efficiency of $\eta \simeq 9(2) \unit{\%}$ for STIRAP, which gives $N_{\mathrm{mol}} = 14.4(3.5) \times 10^3$ molecules produced and a dissociation efficiency of $55(20) \unit{\%}$. The error bars for this number comes from our measurements of all the relevant STIRAP parameters.

\begin{figure}[tb]
\includegraphics[width=\columnwidth]{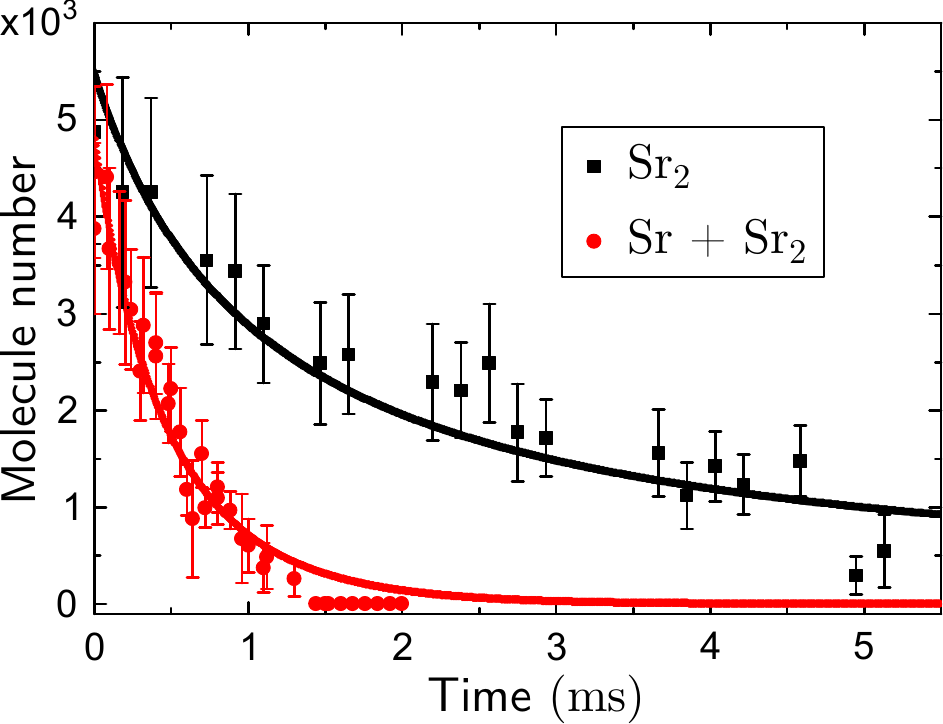}
\caption{\label{fig MoleculeLifetime} (color online) Decay of the number of $\mathrm{Sr}_2$ molecules as a function of hold time, both in a pure sample of molecules (black squares) and in a mixture of atoms and molecules (red circles). The curves are the solutions to the collision rate equations using the fitted rate parameters.}
\end{figure}

We measure the lifetime of the molecules we produced, which originates from inelastic collisions between two molecules or between a molecule and a Sr atom. We measure the molecular lifetime both in a pure $\mathrm{Sr}_{2}$ sample and in a mixture $\mathrm{Sr}_{2} +\mathrm{Sr}$, depending on when the push pulse is applied. The number of molecules as a function of the hold time is shown for both cases in Fig.~\ref{fig MoleculeLifetime}, together with exponential fits. Assuming the STIRAP efficiency provided by our theoretical model, we measure a $1/e$ time $\tau_{\mathrm{Sr_{2}}}=2.7(9) \unit{ms}$ for a pure $\mathrm{Sr}_{2}$ sample containing $5(1) \times 10^3$ molecules, and $\tau_{\mathrm{Sr+Sr_{2}}}=0.54(3) \unit{ms}$ for a mixture of $\simeq 170 \times 10^3$ Sr atoms and $4.5(8)\times 10^3$ molecules. From these lifetimes we can extract collision rate parameters considering losses that arise only from two-body inelastic collisions, when molecules are changing vibrational level and gaining enough kinetic energy to leave the trap \cite{Idziaszek2010UniversalRateCollMol}. Since the decay time is fast compared to the trapping frequencies, we assume the spatial distribution of Sr atoms to be the same as the one before STIRAP, and the $\mathrm{Sr}_{2}$ distribution to coincide with the one describing the center of mass of an atom pair before STIRAP. We first fit the decay curve of the pure molecular sample and retrieve a two-body molecular collision rate parameter $K_{mm} = 3.4_{-1.2}^{+2.3} \times 10^{-10} \unit{cm^3 / s}$. We then fix this parameter and fit the molecule number decay in the mixture of both atoms and molecules. Neglecting the small variation of the atom number during the experiment time, we retrieve the atom/molecule two-body collision rate parameter $K_{ma}= 8.7_{-3.5}^{+5.3} \times 10^{-11} \unit{cm^3 / s}$. The stated uncertainties arise both from the statistical variation on the retrieved atoms number and from the measured uncertainties on the parameters presented in Tab.~\ref{tab ParametersBEC}. 

We can compare our measurements with the universal rate parameters for low-energy s-wave inelastic collisions given in \cite{Idziaszek2010UniversalRateCollMol, Deiglmayr2011InelasticMolAtomCollRate}. This model assumes unit probability of reaction at short range, and infers a universal collision rate parameter $K^{\mathrm{ls}}= 2 g (h / \mu) \overline{a}$, where $g$ is either 1 or 2 for distinguishable or indistinguishable particles, respectively. $\overline{a}$ is the van der Waals length, which for asymptotic dispersion potentials of the form $-C_6/r^6$ is $\overline{a}= 0.47799 \times (2\mu C_6 / \hbar^2)^{1/4}$. The $C_6$ parameter is here the effective $C_6$ derived from the atomic coefficient $C_6^a$, i.e. $C_6= 4 C_6^a$ and $C_6 =2 C_6^a$ for molecule-molecule and molecule-atom collisions, respectively. For Sr this model predicts $K_{ma} =  6.8 \times 10^{-11} \unit{cm^3 s^{-1}}$ and $K_{mm}= 6.0 \times 10^{-11} \unit{cm^3 s^{-1}}$. The former value is consistent with our measurement, while the latter is roughly a factor of 5 smaller that the measured one. This could be due to an underestimation of the molecular density in our sample, or a departure from universality \cite{Idziaszek2010UniversalRateCollMol}.

\section{STIRAP limitations}
\label{sec:STIRAPlimitations}

We now analyse the STIRAP efficiency using the theory discussed in \cite{Drummond2002STIRAPmolBEC, Ciamei2017EffProductionSr2Mol}. Since our system fulfills ${\Omega_{\mathrm{FB}}< \gamma_e}$, we operate in a regime of strong dissipation. The three relevant parameters in the problem are then $\alpha=\tilde{\gamma} T$, $A = \delta / \tilde{\gamma}$, and $T$, where $\tilde{\gamma}=\Omega_{\mathrm{FB,BB}}^2 / \gamma_{e}\approx \Omega_{\mathrm{FB}}^2 / \Gamma_{e}$. We consider the light shift $\delta_{\mathrm{FB}}$ induced by $L_{\mathrm{FB}}$ on the binding energy of state $\vert m \rangle$ as the only contribution to the two-photon detuning $\delta$, therefore neglecting the time-dependent mean field shifts addressed in \cite{Drummond2002STIRAPmolBEC}, which thus leads to $A \approx \delta_{\mathrm{FB}} / \tilde{\gamma}$. Indeed the optimized parameters for STIRAP lead to a shift $\delta_{\mathrm{FB}}=2 \pi \times 135 \unit{kHz}$, which is much bigger than the mean-field shift of about $2 \pi \times 1 \unit{kHz}$, thus justifying this approximation. An efficient STIRAP requires $\alpha \gg \pi^2$, which ensures the adiabaticity of the transfer, $|A| \ll 1$, which ensures a long lifetime of the dark-state superposition, and $T\ll \tau_{\mathrm{Sr+Sr_{2}}}$, which ensures small losses from state $\vert m \rangle$. As discussed in detail in \cite{Ciamei2017EffProductionSr2Mol}, the parameter $A$ depends only on molecular physics and is here given by $\left\vert A \right\vert \approx \left\vert\delta_{\mathrm{FB}} \right\vert / \tilde{\gamma} \approx (\hbar \Omega_{\mathrm{FF}}^{2} / 2 \Delta E_{e}) \, \gamma_e/\Omega_{\mathrm{FB}}^{2} = (\Omega_{\mathrm{FF}} / \Omega_{\mathrm{FB}})^2 \, \hbar \gamma_e /2 \Delta E_{e} = 1 / (G^2 \, N \, \mathrm{FCF_{FB}^2} ) \,\,  \hbar \gamma_e /2 \Delta E_{e}$, where $N$ is the BEC atom number, and $\mathrm{FCF_{FB}}$ and $G=1/ \sqrt{3}$ are respectively the Franck-Condon factor and the geometric factor of the free-bound transition \cite{Machholm2001TheoCollAlkEarth}. From our measurements we derive $|A| \approx 18 \gg 1$ and $\alpha \simeq 4 < \pi^2$. We think the experimental constraint on $\alpha$ is imposed by the strong losses induced by the high $\left\vert A \right\vert$, which prevent a near-unit efficiency STIRAP. We conclude that the main limitation to the STIRAP efficiency in our system is the short lifetime of the dark state, due to the time-dependent light shift $\delta_{\mathrm{FB}}$.

One way to significantly increase the dark-state lifetime is to compensate $\delta_{\mathrm{FB}}$ at all times during the STIRAP sequence. This has been achieved for our $\Lambda$ scheme in the case of STIRAP on Sr atoms in a Mott insulator \cite{Ciamei2017EffProductionSr2Mol}, allowing for a transfer efficiency higher than $80 \unit{\%}$. By numerical simulation of eq.~(\ref{eq BEC Model}), we find that adapting this compensation technique to our case will allow us to reach an efficiency of $40 \unit{\%}$. The remaining limitation will then be the short molecular lifetime, which will put an upper bound on the parameter $\alpha$. However, since $\alpha \propto \mathrm{FCF}^{2}_{\mathrm{FB}}$, the STIRAP efficiency can be further increased by choosing a $\Lambda$ scheme with a stronger free-bound transition, i.e. with a bigger Franck-Condon factor, if available. A STIRAP with near-unit efficiency applied on a BEC would ensure the coherent optical production of a molecular BEC \cite{Jochim2003molBEC, Greiner2003MolBECFromFermi, Yan2013RabiFlopBECMolviaPA}. In particular, the short molecular lifetime, which might prevent the sample from reaching thermal equilibrium, could be circumvented by a second STIRAP towards the potentially much more stable rovibrational groundstate \cite{Ni2008HighPSDMol, Danzl2008BoundMolStirap}.

\section{Conclusion and outlook}
\label{sec:conclusion}

In conclusion, we have demonstrated the bosonic enhancement of the free-bound transition dipole moment in a Sr BEC and we have exploited it to coherently produce ultracold $\mathrm{Sr}_2$ ground-state molecules using a STIRAP pulse sequence. We derive a STIRAP efficiency of $9(2) \unit{\%}$ and find it to be strongly limited by the finite lifetime of the dark-state superposition. However, we calculate that by optically compensating the time-dependent light shifts, as demonstrated in \cite{Ciamei2017EffProductionSr2Mol}, an efficiency of $40 \unit{\%}$ should be achievable. Further increase in the efficiency is possible if stronger free-bound lines are used. We directly observe the products of this \textit{superchemistry} reaction \cite{Heinzen2000Superchemistry} and measure their inelastic collision rate parameters. Inelastic collisions result in lifetimes short but sufficient to allow a second STIRAP toward potentially more stable rovibrational ground-state molecules \cite{Ni2008HighPSDMol, Danzl2008BoundMolStirap}. Provided higher transfer efficiencies and longer lifetimes, it might be possible to produce a molecular sample that would be stable enough to reach thermal equilibrium and would feature a mBEC.

\begin{acknowledgments}

We gratefully acknowledge funding from the European Research Council (ERC) under Project No. 615117 QuantStro. B.P. thanks the NWO for funding through Veni grant No. 680-47-438.

\end{acknowledgments}


\end{document}